\documentclass[prb,reprint,superscriptaddress,showpacs,amsmath,amssymb,floatfix,
]{revtex4-1}
\usepackage{dcolumn}
\usepackage{bm}
\usepackage{turnstile}
\usepackage{graphicx}
\usepackage{amsmath,amssymb}
\usepackage{url}
\usepackage{multirow}
\usepackage{float}
\usepackage[toc,page]{appendix}
\usepackage{color}
\usepackage{ulem}

\begin{document}
\title{Superconducting diode effect in topological hybrid structures}
\author{T.~Karabassov}\email{tkarabasov@hse.ru}
	\affiliation{HSE University, 101000 Moscow, Russia}
\author{E. S. Amirov}
	\affiliation{HSE University, 101000 Moscow, Russia}
\author{I.~V.~Bobkova}
    \affiliation{Institute of Solid State Physics, Chernogolovka, Moscow reg., 142432 Russia}
    \affiliation{Moscow Institute of Physics and Technology, Dolgoprudny, 141700 Russia}
    \affiliation{HSE University, 101000 Moscow, Russia}
\author{A.~A.~Golubov}
	\affiliation{Faculty of Science and Technology and MESA$^+$ Institute for Nanotechnology,
		University of Twente, 7500 AE Enschede, The Netherlands}
\author{E. A. Kazakova}
	\affiliation{Sechenov First Moscow State Medical University, 119991 Moscow, Russia}
\author{A.~S.~Vasenko}
    \affiliation{HSE University, 101000 Moscow, Russia}
    \affiliation{I.E. Tamm Department of Theoretical Physics, P.N. Lebedev Physical Institute, Russian Academy of Sciences, 119991 Moscow, Russia}

\begin{abstract}
Currently, the superconducting diode effect (SDE) is actively discussed due to large application potential in superconducting electronics. In particular, the superconducting hybrid structures based on three-dimensional (3D) topological insulators are among the best candidates due to the strongest spin-orbit coupling (SOC). Most of the theoretical studies of the SDE focus either on full numerical calculation, which is often rather complicated or on the phenomenological approach. In the present paper we perform a comparison of the linearized and nonlinear microscopic approaches in the superconductor/ ferromagnet/ 3D topological insulator (S/F/TI) hybrid structure. Employing the quasiclassical Green's function formalism we solve the problem self-consistently. We show that the results obtained by the linearized approximation are not qualitatively different from the nonlinear solution. Main distinction in the results between the two methods is quantitative, i. e. they yield different supercurrent amplitudes. However, when calculating the so-called diode quality factor the quantitative difference is eliminated and both approaches can result in a good agreement. 
\end{abstract}
\maketitle


\section{Introduction}

\begin{figure}[t]
\includegraphics[width=\columnwidth]{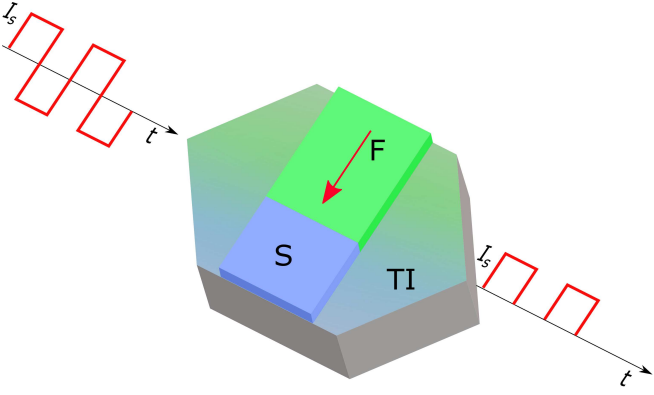}
\caption{Schematic representation of the superconducting diode, where two-dimensional (2D) S/F structure is placed on the surface of a three-dimensional (3D) topological insulator.\label{model}}
\end{figure}

The field of superconducting electronics is an important area of research and development the hybrid quantum devices with lower power consumption. Superconducting hybrid structures consisting of superconductor and non-superconducting material (normal metal N, ferromagnet F, etc.) operate by means of the proximity effect. This effect can be described as a leakage of the superconducting correlations into the adjacent non-superconducting layer \cite{RevBuzdin2005, RevGolubov2004, RevBergeret2005, Demler1997, Ozaeta2012R, Bergeret2013, FuKane2008, Stanescu2010, Black-Schaffer2011,Yano2019,Romano2021}. Superconductor/ferromagnet (S/F) structures were proposed in many nanoelectronic applications like memory devices \cite{Soloviev2017}, quantum and classical logic devices \cite{Soloviev2017,Chernodub2022}, artificial neural networks \cite{Soloviev2018}, detectors and bolometers \cite{Gordeeva2020}, nanorefrigerators \cite{Ozaeta2012, Kawabata2013} and spin-valves \cite{Neilo2022}. Placing the two-dimensional (2D) S/F structures on a surface of a 3D topological insulator, a material with strong spin-orbit coupling, may add a new functionality and create the so called superconducting diode, see Fig.~\ref{model}.

The superconducting diode effect (SDE) is an active area of reseach because of a great application potential in the field of superconducting electronics and spintronics. Generally, the SDE is observed in the two-dimensional superconducting systems with broken inversion and time reversal symmetries \cite{Nadeem_arxiv}. While the former usually implies the presence of the spin-orbit field, the latter can be achieved by the exchange field from the ferromagnet or by exposing the system to an external magnetic field. Big advance has been made since the experimental discovery of the diode effect by Ando \textit{et al} \cite{Ando2020}. There have been numerous reports on both experimental \cite{Ando2020,Bauriedl_arxiv,Shin_arxiv,Trahms_arxiv} and theoretical studies \cite{Daido2022,He_arxiv,Yuan_arxiv,Scammell_arxiv,Ilic_arxiv,Devizorova2021,dePicoli_arxiv} of the SDE. The hybrid SDE devices deserve special attention \cite{Devizorova2021, Kokkeler2022, Karabassov2022}. In such structures the ingredients for the SDE effect are brought together by the proximity effect. For instance, S/F/TI hybrid structure is a promissing platform for realization of the superconducting diode \cite{Karabassov2022}. It should be noticed that placing the S/F structures on the surface of a 3D topological insulator leads to a number of striking phenomena of magnetoelectric nature \cite{Bobkova2004,Mironov2017,Pershoguba2015,Malshukov2020,Malshukov2020_2,Malshukov2020_3}. Moreover, new electronic states have been predicted to appear in such structures including magnetic monopoles \cite{Qi2009} and Majorana fermions \cite{Tanaka2009,Maiellaro2021,Mazziotti2018,Maiellaro2022}. It has been also predicted that the presence of the helical magnetization in the F layer leads to the nonmonotonic dependence of the critical temperature on the F layer width in S/F/TI structures \cite{Karabassov2021}.

Majority of the existing theoretical studies on the SDE focus either on the microscopic numerical calculations \cite{Yuan_arxiv,Ilic_arxiv,Legg2022_arxiv,Davydova_arxiv} or on the phenomenological approach \cite{Devizorova2021,He_arxiv}. In this work we consider both linear and nonlinear approaches to calculate the SDE in the hybrid S/F/TI structure. We use the microscopic quasiclassical Green's functions formalism in the diffusive regime. We provide the comparison between the results obtained by linear and nonlinear methods and discuss their ranges of applicability.

%

\section{Materials and Methods}
In this section we present the model under consideration.  The system is described by the following model Hamiltonian:
\begin{equation}
H= H_0 + H_F + H_{S},
\end{equation}
where
\begin{eqnarray}
H_0=\int d^2 r \Psi^\dagger (\bm r)\bigl[-i\alpha(\bm \nabla_{\bm r}\times \hat z)\bm \sigma - \mu +  V(\bm r) \bigr]\Psi(\bm r), \\
H_F = - \int d^2 r \Psi^\dagger (\bm r) \bigl[ \bm h \bm \sigma \bigr] \Psi(\bm r), \\
H_S = \Delta(\bm r)\Psi^\dagger_\uparrow (\bm r) \Psi^\dagger_\downarrow (\bm r) + \Delta^*(\bm r)\Psi_\downarrow (\bm r) \Psi_\uparrow (\bm r).
\end{eqnarray}
Here $\Psi^\dagger(\bm r)=(\Psi^\dagger_\uparrow(\bm r),\Psi^\dagger_\downarrow(\bm r))$ is the creation operator of an electron at the 3D TI surface, $\hat z$ is the unit vector normal to the surface of TI, $\alpha$ is the Fermi velocity of electrons at the 3D TI surface and $\mu$ is the chemical potential. $\bm \sigma = (\sigma_x, \sigma_y, \sigma_z)$ is a vector of Pauli matrices in spin space and $\bm h = (h_x, h_y, 0)$ is an in-plane exchange field, which is assumed to be nonzero only at $x<0$. The superconducting pairing potential $\Delta $ is nonzero only  at $x>0$. Therefore, effectively the TI surface states are divided into two parts: one of them at $x<0$ possesses $h \neq 0$ and can be called "ferromagnetic", while the other part corresponding to $x>0$ with $\Delta \neq 0$ can be called "superconducting". Below we will use subscripts $f$ and $s$ to denote quantities, related to the appropriate parts of the TI surface. The potential term $V(\bm r)$ includes the nonmagnetic impurity scattering potential $V_{imp}=\sum \limits_{\bm r_i}V_i \delta(\bm r - \bm r_i)$, which is of a Gaussian form $\langle V(\bm r)V(\bm r')\rangle = (1/\pi \nu \tau)\delta(\bm r - \bm r')$ with $\nu=\mu/(2\pi \alpha^2)$, and also possible interface potential $V_{int}(\bm r) = V\delta(x)$. 

The superconductivity and in-plane exchange field is assumed to be proximity induced due to adjacent superconducting and ferromagnetic layers. Thus we can imagine the system to be a planar hybrid structure that consists of superconductor S and ferromagnetic layer F on top of three-dimensional topological insulator TI  as shown schematically in Fig.~\ref{model}. The role of the TI surface is to provide a strong spin-orbit coupling which produces a full spin-momentum locking effect. In this case only one helical band which crossing the Fermi energy is present. We employ the quasiclassical Green's function formalism in the diffusive regime. In principle Green's function matrices have two degrees of freedom that are particle-hole and spin. In our model the spin structure is characterized by a projector onto the conduction band:
\begin{equation}
\check g_{s,f}(\bm n_F, \bm r, \varepsilon)= \hat g_{s,f}(\bm r, \varepsilon)\frac{(1+\bm n_\perp \bm \sigma)}{2},
\label{spin_structure}
\end{equation}
where $\hat g_{s(f)}$ is the spinless Green's functions matrix in the particle-hole space in the superconducting (ferromagnetic) part of the 3D TI layer, $\bm n_F=\bm p_F/p_F=(n_{F,x},n_{F,y},0)$ is a unit vector directed along the quasiparticle trajectory and $\bm n_\perp=(n_{F,y},-n_{F,x},0)$ is a unit vector perpendicular to the quasiparticle trajectory and directed along the quasiparticle spin, which is locked to the quasiparticle momentum. 

In our theoretical analysis, we consider the diffusive limit,
in which the superconducting coherence length is given by
expression $\xi_{s} = \sqrt{D_{s}/ 2 \pi T_{cs}}$, where $D_{s}$ is the diffusion coefficient and $T_{cs}$ is the critical temperature of the bulk superconductor (we assume $\hbar = k_B = 1$) and the elastic scattering length $\ell \ll \xi_{s}$. We also neglect the nonequilibrium effects
in the structure \cite{VH, Arutyunov2011, Arutyunov2018}.

In the following we outline the nonlinear and linear equations to calculate the SDE effect in the system under consideration.

\subsection{Nonlinear Usadel equations}

The quasiclassical Usadel equation for spinless Green's functions is\cite{Zyuzin2016,Bobkova2017}
\begin{equation}\label{Usadel_general}
D \hat{\nabla}\left(\hat{g} \hat{\nabla} \hat{g} \right)= \left[\omega_n \tau_z + i \hat{\Delta}, \hat{g}\right].
\end{equation}
Here $D$ is the diffusion constant, $\tau_z$ is the Pauli matrix in the particle-hole space, $\hat{\nabla} X = \nabla X + i \left(h_x \hat{e}_y - h_y \hat{e}_x\right) \left[\tau_z, \hat{g}\right]/\alpha$. The gap matrix $\hat{\Delta}$ is defined as $\hat{\Delta}= \hat{U} i \tau_x \Delta (x) \hat{U}^\dagger$, where $\Delta(x)$ is a real function and transformation matrix $\hat{U}= \exp \left( i q y \tau_z/2 \right)$ . The finite center of mass momentum $q$ takes into account the helical state. The Green's function matrix is also transformed as $\hat{g}= \hat{U} \hat{g}_q \hat{U}^\dagger$.
To facilitate the solution procedures of the nonlinear Usadel equations we employ $\theta$ parametrization of the Green's functions\cite{Belzig1999},
\begin{equation}
\hat{g}_q= 
\begin{pmatrix}
\cos \theta & \sin \theta \\
\sin \theta & -\cos \theta
\end{pmatrix}.
\end{equation}
Substituting the above matrix into the Usadel equation \eqref{Usadel_general}, we obtain in the S part of the TI surface $x>0$:
\begin{equation}
\xi_{s}^2 \pi T_{cs} \left[ \partial_x^2 \theta_{s} - \frac{q^2 }{2} \sin 2 \theta_s \right] =\omega_n \sin{\theta_{s}} - \Delta (x) \cos{\theta_{s}}, \nonumber
\end{equation}
and in the F part $x<0$:
\begin{equation}
\xi_{f}^2 \pi T_{cs}  \left[ \partial_x^2 \theta_{f} - \frac{q_m^2 }{2} \sin 2 \theta_f \right] =\omega_n \sin{\theta_{f}},
\end{equation}
where $\xi_{f} = \sqrt{D_{f}/ 2 \pi T_{cs}}$, and $D_f$ is the diffusion coefficient of the ferromagnetic layer. $q_m = q + 2 h /\alpha$ and $X_{s(f)}$ means the value of $X$ in the S(F) part of the TI surface, respectively. The self-consistency equation for the pair potential reads,
\begin{equation}
\Delta (x) \ln \frac{T_{cs}}{T} = \pi T \sum_{\omega_n} \left( \frac{\Delta (x)}{|\omega_n|} - 2 \sin \theta_s \right).
\end{equation}
We supplement the above equations with the following boundary conditions at the S/F interface ($x=0)$ \cite{KL},
\begin{eqnarray}\label{BC}
\gamma_B \frac{\partial \theta_f }{\partial x}\Big\vert_{x=0} = \sin \left( \theta_s - \theta_f \right),\\
\frac{\gamma_B}{\gamma} \frac{\partial \theta_s}{\partial x}\Big\vert_{x=0} = \sin \left( \theta_s - \theta_f \right),
\end{eqnarray}
where $\gamma = \xi_s \sigma_f/ \xi_f \sigma_s$, $\gamma_B = R \sigma_f / xi_f$, and $\sigma_{s(f)}$ is the conductivity of the S (F)layer. 
The parameter $\gamma$ determines the strength
of suppression of superconductivity in the S lead near
the interface compared to the bulk: no suppression occurs
for $\gamma = 0$, while strong suppression takes place for $\gamma \gg 1$.
The parameter $\gamma_B$ is the dimensionless parameter, describing the 
transparency of the S/F interface \cite{KL, VB1, VB2}.
To complete the boundary problem we also set boundary
conditions at free edges,
\begin{equation}\label{BC_vac}
\frac{\partial \theta_f }{\partial x}\Big\vert_{x=-d_f}=0, \quad \frac{\partial \theta_s }{\partial x}\Big\vert_{x=d_s}=0.
\end{equation}

In order to calculate the superconducting current we utilize the expression for the supercurrent density
\begin{equation}
\textbf{J}_{s(f)}= \frac{- i \pi \sigma_{s(f)}}{4 e} T \sum_{\omega_n} Tr \left[ \tau_z \hat{g}_{s(f)} \hat{\nabla} \hat{g}_{s(f)} \right].
\end{equation}
Performing the unitary transformation $U$, the current density transforms as follows:
\begin{eqnarray}
{j}_y^s (x)=- \frac{\pi \sigma_s q }{2 e} T \sum_{\omega_n} \sin^2 \theta_s, \\
{j}_y^f (x)=- \frac{\pi \sigma_n }{2 e} \left[ q  + \frac{2 h}{\alpha}\right] T  \sum_{\omega_n} \sin^2 \theta_f.
\end{eqnarray}
The total supercurrent flowing via the system along the $y$-direction can be calculated by integrated the current density of the total width of the S/F bilayer $d_f+d_s$:
\begin{equation}\label{I_total}
I= \int_{-d_f}^{0} {j}^f_y (x) dx  + \int_{0}^{d_s} {j}^s_y (x) dx .
\end{equation}
%
\subsection{Linear Usadel equations}
In the limit when $T \approx T_{c}$, the Usadel equations \eqref{Usadel_general} can be linearized, since the normal Green's function is close to unity, i. e.  $\hat{g}_q \approx \tau_z + \theta\left(x\right) \tau_x$. 

In the superconducting S layer ($0 < x <d_s$) the linearized Usadel equation for the spinless amplitude $\theta_s$ reads\cite{Belzig1999,Usadel,Zyuzin2016,Bobkova2017}
\begin{equation}\label{Usadel_S_q}
\xi_s^2 \pi T_{cs} \left(\partial_x^2 -q^2\right) \theta_s - \omega_n \theta_s + \Delta = 0.
\end{equation}
In the ferromagnetic region of the TI the linearized Usadel equation takes the form
\begin{equation} \label{Usadel_TIq}
\partial_x^2 \theta_f= \left[\frac{{\omega_n}}{\xi_f^2 \pi T_{cs}} +  q_m^2\right]\theta_f.
\end{equation}
The solution of Eq.~\eqref{Usadel_TIq} can be found in the form
\begin{equation}\label{f_T}
\theta_f= C(\omega_n) \cosh k_{q} \left( x + d_f \right),
\end{equation}
where
\begin{align}\label{kq}
k_{q} = \sqrt{\frac{|{\omega_n}|}{\xi_f^2 \pi T_{cs}} + q_m^2}.
\end{align}
Here $C(\omega_n)$ is to be found from the boundary conditions.
Using boundary conditions \eqref{BC} we can write the problem in a closed form with respect to the Green function $f_s$. At $x=0$ the boundary conditions can be written as:
\begin{equation}\label{boundary}
\xi_s \frac{\partial \theta_s(0)}{\partial x} = W^q(\omega_n) \theta_s(0),
\end{equation}
where,
\begin{equation}
W^q(\omega_n)= \frac{\gamma}{\gamma_B + A_{qT} (\omega_n)}, \quad
A_{qT}(\omega_n)= \frac{1}{k_{q} \xi_f} \coth{ k_{q} d_f}.
\end{equation}
In general the boundary condition \eqref{boundary} can be complex. But in the considered system $A_{qT}$ is real. Hence the condition \eqref{boundary} coincides with its real-valued form.

Then we write the self-consistency equation for $\Delta$ considering only positive Matsubara frequencies,
\begin{equation}\label{Delta+}
\Delta \ln \frac{T_{cs}}{T} = \pi T \sum_{\omega_n > 0} \left ( \frac{2 \Delta} {\omega_n} - 2 \theta_s \right),
\end{equation}
as well as the Usadel equation in the superconducting part,
\begin{equation}\label{finUsadelS}
\xi_s^2 \left(\frac{\partial^2 \theta_s}{\partial x^2} - \kappa_{qs}^2 \theta_s\right) + \frac{\Delta}{\pi T_{cs}} = 0.
\end{equation}
Within the linearized Usadel equations the supercurrent is also calculated in the self-consistent manner using Eqs. \eqref{boundary} - \eqref{finUsadelS}.

\subsubsection{Single-mode approximation}
In the framework of the so-called single-mode approximation the solution in S is introduced in the form\cite{Fominov2002,Karabassov2019},
\begin{equation}\label{Fssingle}
\theta_s(x,\omega_n)=f(\omega_n) \cos\left(\Omega\frac{x-d_s}{\xi_s}\right),
\end{equation}
\begin{equation} \label{Dsingle}
\Delta(x)=\delta \cos \left(\Omega\frac{x-d_s}{\xi_s}\right).
\end{equation}
The solution presented above automatically satisfies boundary condition \eqref{BC_vac} at $x=d_s$.
Substituting expressions \eqref{Fssingle} and \eqref{Dsingle} into the Usadel equation for $\theta_s$ \eqref{finUsadelS} yields
\begin{align}\label{f_om}
f(\omega_n)=\frac{\delta}{\omega_n + \Omega^2 \pi T_{cs} +q^2 \xi_s^2 \pi T_{cs}}.
\end{align}
Employing the single-mode approximation and using the solution in TI layer as well as the boundary conditions we can find the total supercurrent flowing through the system in $y$ direction. We obtain the following expression
\begin{eqnarray}
I=- \frac{\pi}{4 e}  T  \sum_{\omega_n} f^2(\omega_n)  \left[ \sigma_s q C_s +  \frac{\sigma_n  }{\beta ^2} \cos^2 \left(\Omega \frac{d_s}{\xi_s}\right) q_m C_f\right],  \\
C_s=\left(d_s + \frac{\xi_s}{2 \Omega} \sin \left(2 \Omega \frac{d_s}{\xi}\right)\right), \nonumber \\
C_f = \left( d_f + \frac{1}{2 k_q} \sinh \left(2 k_q d_f\right)\right). \nonumber
\label{current_final}
\end{eqnarray}
In Eq.~(\ref{current_final}) coefficient $\beta$ is defined as, 
 \begin{align}
 \beta = \gamma_B  k_q \xi_f \sinh k_q d_f + \cosh k_q d_f, 
 \end{align}
 and $\Omega$ is calculated from the boundary condition for the single-mode approximation \eqref{boundary}.
In the following section we present the results of supercurrent calculation for both linear (self-consistent and single-mode) and nonlinear approaches.

\section{Results}
In this section we present the results of the calculations based on the model presented above. For simplicity we set $\xi_s = \xi_f= \xi$.

\begin{figure}[H] 
\includegraphics[width=\columnwidth]{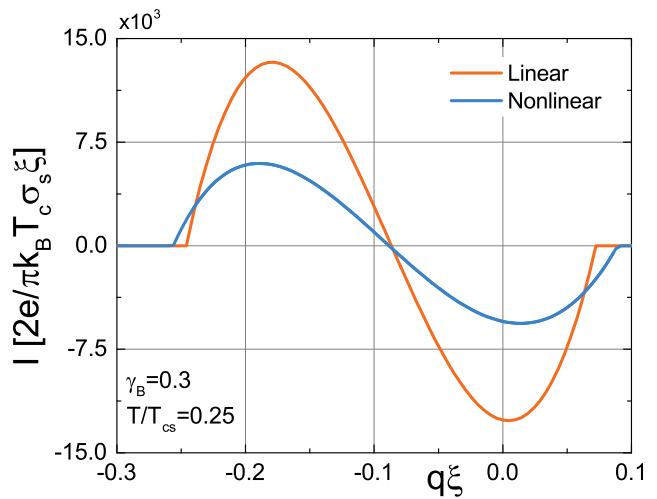}
\caption{ The total supercurrent $I$ as a function of the Cooper pair momentum $q$ calculated self-consistently via linear and nonlinear methods. The parameters of the calculation: $d_s =1.2 \xi, d_f = \xi ,\gamma =0.5 , \xi h/\alpha =0.3$.\label{fig1}}
\end{figure}

In Fig. \ref{fig1} we compare $I(q)$ dependencies calculated by linear and nonlinear approaches. Both of these curves were calculated in a numerical self-consistent approach. Firstly we can notice that linearised solution results in higher values of the critical currents. As expected the linearised approach does not capture nonlinearities in the current behavior as a function of $q$.

\begin{figure}[H]
\includegraphics[width=\columnwidth]{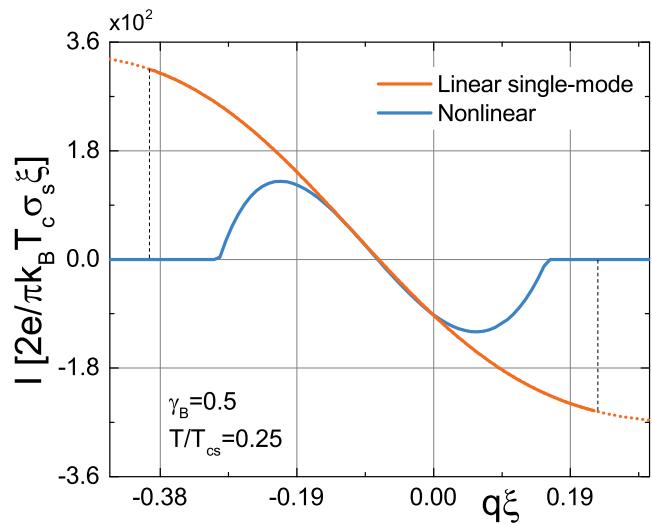}
\caption{The total supercurrent $I$ as a function of the Cooper pair momentum $q$ calculated self-consistently via linear sigle-mode approximation and self-consistent nonlinear method. Vertical dotted line corresponds to the critical temperature calculated by the single-mode approximation. The parameters of the calculation: $d_s =1.2 \xi, d_f = \xi ,\gamma =0.5 , \xi h/\alpha =0.3$.\label{fig2}}
\end{figure}

When calculating the critical temperature in the hybrid structure it is common to use the single-mode aproach within the linearized Usadel equations. We test the application possibility of the single-mode method to calculate the supercurrent. The main disadvantages of Eqs.~\eqref{Fssingle} - \eqref{Dsingle} are that these expressions disregard the dependencies of the amplitude $\delta$ on the parameter $q$. Moreover the amplitude of the pair potential can not be obtained within the solution provided by the single-mode, i.e. $\delta$ remains as a fitting parameter. In fig.~\ref{fig2} we provide comparison between the full nonlinear approach and the single-mode approximation. We observe that the supercurrent derived by the single-mode can be in a fairly good agreement with the nonlinear method in the vicinity of the equilibrium value of $q=q_0$. However for larger values of $q-q_0$ it is clear that the single-mode approach tends to fail resulting in a much larger values of the critical current.

It is more instructive to discuss the diode quality factor, which is defined in the following way
\begin{equation}
\eta = \frac{\Delta I_c}{I_c^+ + |I_c^-|} = \frac{I_c^+ - |I_c^-|}{I_c^+ + |I_c^-|},
\end{equation}
where $I_c^{+ (-)}$ is the critical supercurrent for positive (negative) direction.

\begin{figure}[H]
\includegraphics[width=\columnwidth]{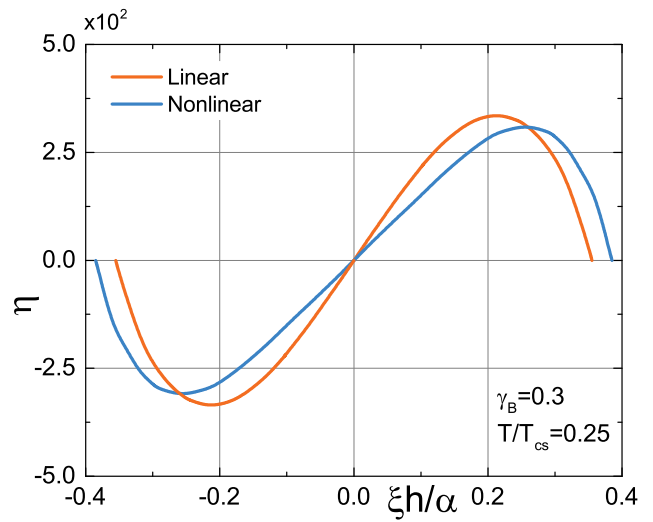}
\caption{ SDE quality factor $\eta$ as a function of exchange field $h$. The parameters of the calculation: $d_s =1.2 \xi, d_f = \xi ,\gamma =0.5$. \label{fig3}}
\end{figure}

In Fig.~\ref{fig3} we demonstrate the SDE quality factor as a function of exchange field $\xi h/ \alpha$ for nonlinear and linear approaches. We can emphasize that the qualtity factors calculated for both cases are quite similar despite the fact that $I(q)$ may be substantially different both quantitatively and qualitatively (Fig. \ref{fig1}). This fact can be connected with the definition of the quality factor $\eta$. Namely since $\eta$ is defined as a ratio between a sum and a difference of the critical currents it loses the information about the current values and $q$ dependencies.
%
\section{Discussion}
In this work we have calculated the superconducting diode effect using three different ways, which include linear and nonlinear equations.The results obtained above may suggest several conclusions. The simplified single-mode approximation for the linearized Usadel equation is only applicable for a qualitative critical temperature calculation. Although single-mode approach may be used at the vicinity of $(q - q_0)$, it does not capture possible $q$ dependency of the pair potential and fails at larger $|q - q_0|$. When operating close to the critical temperature full solution of the linearized Usadel equation gives adequate results. Particularly $\eta$ calculated via the linearized approach can be in a good agreement with the nonlinear case (Fig. \ref{fig3}). Nevertheless, in order to get a valid description of the helical state and the SDE in a wide range of parameters one should use fully nonlinear equations.

\section{Acknowledgements}
T.K. acknowledges the financial support by the Foundation for the Advancement of Theoretical Physics and Mathematics “BASIS” grant number 22-1-5-105-1. A.S.V. acknowledge financial support from the Mirror Laboratories Project and the Basic Research Program of the HSE University.

\bibliography{diode.bib}

\end{document}